\documentclass[reprint,aps,prc,amsmath,amssymb,nofootinbib,superscriptaddress]{revtex4-2}

\usepackage{mathtools}
\usepackage{braket} 
\usepackage[dvipsnames]{xcolor}  
\usepackage{float}
\usepackage[T1]{fontenc}
\usepackage[utf8]{inputenc}
\usepackage{graphicx,color,rotating,pifont}
\usepackage{amsmath,amssymb}
\usepackage{mathtools}
\usepackage{subfigure}
\usepackage{siunitx}
\usepackage{bm}
\usepackage{ae}
\usepackage{dcolumn}
\usepackage{txfonts}
\usepackage{tensor}
\usepackage{braket}
\usepackage{booktabs}
\usepackage{xspace}
\usepackage{soul}
\usepackage{multirow}
\setstcolor{red}
\setul{}{.2ex} 

\usepackage[normalem]{ulem}

\usepackage{tikz}
\usetikzlibrary{positioning,calc,shapes,decorations.markings,decorations.pathmorphing}
\tikzset{asg/.cd,
  omega-vertex/.style={circle,solid,draw=black,fill=white,minimum size=5pt, inner sep=0pt},
  dbd-vertex/.style={coordinate},
  pline/.style={thick, postaction={decorate}, decoration={markings, mark=at position .5 with {\arrow[xshift=2pt]{stealth}}}},
  hline/.style={thick, postaction={decorate}, decoration={markings, mark=at position .5 with {\arrowreversed[xshift=-2pt]{stealth}}}},
  shift arrow/.style={/pgf/decoration/transform={xshift=#1}},
  shift arrow/.default=-2pt,
  dbd-2b/.style={decorate, decoration=snake},
  omega-2b/.style={densely dashed},
  neutron/.style={draw=blue},
  proton/.style={draw=red},
}

\usepackage[colorlinks,linkcolor=blue,anchorcolor=blue,citecolor=blue]{hyperref} 
\hypersetup{
   colorlinks=true, 
   linkcolor=blue,  
   citecolor=blue,  
   filecolor=blue,  
   urlcolor=blue    
} 

\DeclareMathSymbol{\NS}{\mathord}{AMSb}{"4E}

\DeclareSIUnit{\fm}{\femto\meter}

\newcommand{\beq}{\begin{equation}}
\newcommand{\eeq}{\end{equation}}
\newcommand{\beqn}{\begin{eqnarray}}
\newcommand{\eeqn}{\end{eqnarray}}
\newcommand{\bsub}{\begin{subequations}}
\newcommand{\esub}{\end{subequations}}
\newcommand{\bpm}{\begin{pmatrix}}
\newcommand{\epm}{\end{pmatrix}}

\newcommand\identity{1\kern-0.25em\text{l}}

\begin{document}

\title{Impact of isovector pairing fluctuation on neutrinoless double-beta decay in multi-reference covariant density functional theory}

  \author{C. R. Ding}    
  \affiliation{School of Physics and Astronomy, Sun Yat-sen University, Zhuhai 519082, P.R. China}  

  \author{X. Zhang}    
  \affiliation{School of Physics and Astronomy, Sun Yat-sen University, Zhuhai 519082, P.R. China}  
  
  \author{J. M. Yao}   
 \email{Corresponding author: yaojm8@sysu.edu.cn}
  \affiliation{School of Physics and Astronomy, Sun Yat-sen University, Zhuhai 519082, P.R. China}  

\author{P. Ring}
 \affiliation{Physik Department, Technische Universit\"{a}t M\"{u}nchen, D-85748 Garching, Germany} 

\author{J. Meng}
 \affiliation{State Key Laboratory of Nuclear Physics and Technology, School of Physics, Peking University, Beijing 100871, China}
 \affiliation{Yukawa Institute for Theoretical Physics, Kyoto University, Kyoto 606-8502, Japan}

\date{\today}

\begin{abstract} 

 We extend the multi-reference covariant density functional theory (MR-CDFT) by including fluctuations in quadrupole deformations and average isovector pairing gaps simultaneously for the nuclear matrix elements (NMEs) of neutrinoless double-beta $(0\nu\beta\beta)$ decay in the candidate nuclei \nuclide[76]Ge, \nuclide[82]Se, \nuclide[100]Mo,  \nuclide[130]Te, and \nuclide[136]Xe  assuming the exchange of either light or heavy neutrinos. The results indicate a linear correlation between the predicted NMEs and the isovector pairing strengths, as well as the excitation energies of $2^{+}_1$ and $4^{+}_1$ states. By adjusting the pairing strengths based on the excitation energies of the $2^{+}_1$ states, we calculate the NMEs for $0\nu\beta\beta$ decay, which are reduced by approximately $12\%$ to $62\%$ compared to the results obtained in the previous studies by Song {\em et al.} [Phys. Rev. C95, 024305 (2017)]. Additionally, upon introducing the average isovector pairing gap as an additional generator coordinate in the calculation, the NMEs increase by a factor ranging from $56\%$ to $218\%$.

\end{abstract}
  
\maketitle

\section{Introduction}

Neutrinoless double-beta ($0\nu\beta\beta$) decay is a hypothetical second-order weak-interaction process in which an even-even nucleus decays into its neighbouring even-even nucleus  with two fewer neutrons and two more protons, with the emission of only two electrons~\cite{Furry:1939}. The observation of this process would provide direct evidence for the existence of lepton-number violation processes in nature and implies the existence of a Majorana mass term for the neutrino~\cite{Schechter:1982PRD}. As a result, the search for $0\nu\beta\beta$ decay in atomic nuclei has been a significant research frontier in particle and nuclear physics~\cite{Song:2014,Dolinski:2019,
Yao:2021Review,Cirigliano:2022JPG,Cirigliano:2022Snowmass,Agostini:2022RMP,Adams:2022White_Paper}. However, to date, no signal has been observed. The most sensitive half-life limits have been obtained from experiments on $\nuclide[136]{Xe}$, with $T^{0\nu}_{1/2}>2.3\times 10^{26}$ yr at 90\% C.L.~\cite{Abe:2023PRL}, and on $\nuclide[76]{Ge}$, with $T^{0\nu}_{1/2}>1.8\times 10^{26}$ yr at 90\% C.L.~\cite{GERDA:2020}.

If $0\nu\beta\beta$ decay is mainly driven by the mechanism of exchanging light Majorana neutrinos, the half-life of $0\nu\beta\beta$ decay offers a way to determine the absolute masses of neutrinos, provided that the nuclear matrix element (NME) $M^{0\nu}$ is known. The NME cannot be measured experimentally but relies on nuclear model calculations. However, various nuclear models predict NMEs that differ from each other by a factor of about three or even more~\cite{Menendez:2009,Rodriguez:2010,Barea:2013,Mustonen:2013,Holt:2013,Kwiatkowski:2014,Song:2014,Yao:2015,Hyvarinen:2015,Horoi:2016,Song:2017,Jiao:2017,Yoshinaga:2018,Fang:2018,Rath:2019,Terasaki:2019,Coraggio:2020,Deppisch:2020ztt,Wang:2021,Coraggio:2022PRC,Wang:2023}, causing a large uncertainty in the extracted effective Majorana neutrino mass $\langle m_{\beta\beta}\rangle$ from the half-life of $0\nu\beta\beta$ decay. For $\nuclide[136]{Xe}$, the upper limit of $\langle m_{\beta\beta}\rangle =[36, 156]$ meV is obtained from the most recent measurement on the half-life~\cite{Abe:2023PRL}, where the uncertainty of a factor about $2-3$ comes from the NMEs from different model calculations.  The discrepancy is mainly from the systematic uncertainty, which is however challenging to decrease because different nuclear models have different choices of model spaces~\cite{Engel:2017}.   Therefore, understanding and reducing the discrepancy in the NMEs among different models is of particular importance to provide a strict constraint on neutrino masses and has attracted lots of attention in nuclear theory community~\cite{Yao:2021Review,Agostini:2022RMP}.

Pairing correlation is one of the most important correlations that have a significant impact on the NMEs of $0\nu\beta\beta$ decay. The observed large discrepancy in the NMEs among different nuclear models, except for the recent ab initio studies starting from nuclear chiral forces~\cite{Yao:2020PRL,Belley2021PRL,Novario:2021PRL}, might partially originate from different treatments of pairing correlations in candidate nuclei~\cite{Caurier:2008PRL,Simkovic:2008}. The NME is very sensitive to the strength parameter of pairing forces between nucleons. The competition between the components of two decaying neutrons with the coupled angular momentum $J = 0$ and $J\neq 0$ leads to almost complete cancellation of the contribution to the NME at the long distance~\cite{Simkovic:2008}.  This competition enhances the sensitivity of the final NMEs to the strengths of pairing forces. 
As shown in the interacting shell model (ISM) study based on a schematic pairing plus quadrupole interaction~\cite{Yoshinaga:2018,Higashiyama:2020}, where the pairing strength is treated as a variable parameter, the use of an increased monopole isovector pairing strength by 30\% enhances the excitation energy of the $2^+$ state by 65\%  and the NMEs by 80\% (50\%) for the $0\nu\beta\beta$ decay from $\nuclide[130]{Te}$ to $\nuclide[130]{Xe}$ ($\nuclide[136]{Xe}$ to $\nuclide[136]{Ba}$)~\cite{Higashiyama:2020}. It is worth noting that the NMEs calculated by quasiparticle random-phase approximation (QPRA) methods are highly sensitive to the ratio of isoscalar pairing ($T=0$) strength to isovector pairing ($T=1$) strength, particularly around the value required to reproduce the NME of two-neutrino double beta decay. Even a slight shift in the ratio value can result in a significant change in the predicted NME of $0\nu\beta\beta$ decay. This deficiency can be partially remedied in the renormalized QRPA~\cite{Toivanen:1995PRL,Toivanen:1997PRC,Faessler:1998JPG}, where the  isovector pairing strength is usually adjusted to the NME of $2\nu\beta\beta$ decay and its effect mainly modifies the Fermi matrix element~\cite{Simkovic:2013}. A recent study with a self-consistent QRPA method has shown that a large discrepancy exists in the NMEs from the calculations with a volume and surface types of pairing forces~\cite{Lv:2023}, even though the pairing strengths in the two calculations are optimized to the same pairing gaps from the data of odd-even mass difference. Thus, a more elaborate treatment of pairing correlation between nucleons is required in the studies of the NMEs of $0\nu\beta\beta$ decay.  

The generator coordinate method (GCM)~\cite{Hill:1953,Wong:1975,Ring:1980} has proven to be a powerful tool for nuclear low-lying states~\cite{Bender:2003RMP,Niksic:2011PPNP,Sheikh:2019JPG}, and  it has been applied to calculate NMEs of $0\nu\beta\beta$ decay based on different Hamiltonians and energy density functionals (EDFs)~\cite{Engel:2017,Yao:2021Review}. In the GCM study based on a non-relativistic Gogny force~\cite{Vaquero:2013}, the impact of isovector pairing fluctuation on the NMEs of candidate nuclei was considered by including pairing amplitudes as one of the generator coordinates. It was found that the NMEs increase by a factor of $10\%$–$40\%$. Multi-reference covariant density functional theory (MR-CDFT)~\cite{Yao:2009PRC,Yao:2010}, a combination of GCM with CDFT~\cite{Vretenar:2005PR,Meng:2005PPNP,Meng:2016Book}, has been successfully employed to explore various phenomena concerning nuclear low-lying states~\cite{Niksic:2011PPNP,Yao:2014PRC,Sheikh:2019JPG,Yao:2022HandBook}. This framework provides a beyond relativistic mean-field (RMF) description for the NMEs of $0\nu\beta\beta$ decay~\cite{Song:2014, Yao:2015, Yao:2016PRC,Song:2017}, where the fluctuation in quadrupole shapes has been considered.  In recent years the ab initio version of GCM, called in-medium (IM) GCM~\cite{Yao:2018wq,Yao:2020PRL}, has been developed for nuclear low-lying states and NMEs of $0\nu\beta\beta$ decay in the lightest candidate nucleus starting from nuclear forces from chiral effective field theory and transition operators evolved from multi-reference similarity renormalization group method~\cite{Hergert:2016jk}. The IM-GCM has demonstrated success in studying low-lying states of deformed nuclei. However, its application to heavy candidate nuclei of $0\nu\beta\beta$ decay remains a significant challenge. 

Considering the aforementioned points, this study further extends the framework of MR-CDFT for the  NMEs of $0\nu\beta\beta$ decay. It incorporates fluctuations in both quadrupole shapes and isovector pairing gaps for the five well-established candidate nuclei \nuclide[76]Ge, \nuclide[82]Se, \nuclide[100]Mo, \nuclide[130]Te, and \nuclide[136]Xe assuming the exchange of either light ($\nu$) or heavy ($N$) neutrinos. Since the NMEs are sensitive to the quadrupole deformations and average isovector pairing gaps of the mean-field configurations, taking both of them as generate coordinates will eliminate greatly the sensitivity of the predicted NMEs to the choice of their specific values.  Besides, this work presents the first study of utilizing nuclear low-lying state information for the calibration of NMEs in $0\nu\beta\beta$ decay within an EDF-based GCM framework.
 
The article is organized as follows. In Sec.~\ref{sec:formalism}, we present an introduction to the extended MR-CDFT with fluctuations in both quadrupole shapes and isovector pairing gaps, as well as the formulas for the NMEs of $0\nu\beta\beta$ decay in the mechanisms of exchanging either light or heavy neutrinos.   In Sec.~\ref{sec:results}, we present the result on the correlation relation between the NMEs of $0\nu\beta\beta$ decay, excitation energies of $2^+_1$ and $4^+_1$ states and isovector pairing strengths with the MR-CDFT. Using the pairing strengths adjusted to the excitation energies of $2^+_1$ states, we calculate the NMEs of $0\nu\beta\beta$ decay with and without the isovector pairing fluctuation. A summary and perspective are given in Sec.~\ref{sec:summary}.

\section{Formalism} 
\label{sec:formalism}

\subsection{The MR-CDFT for nuclear low-lying states}

In the MR-CDFT, wave function of nuclear low-lying state is constructed as a superposition of quantum-number projected mean-field wave functions within the GCM~\cite{Ring:1980},
 \begin{equation}
 \label{eq:gcmwf}
 \vert  \Psi^{JMNZ}_\sigma\rangle
 =\sum_{\mathbf{q}} f^{J}_\sigma(\mathbf{q}) \ket{JMNZ, \mathbf{q}},
 \end{equation}
 where $\sigma$ distinguishes different states with the same quantum numbers $JM$. The basis function is constructed as
 \begin{equation}
   \ket{JMNZ, \mathbf{q}} \equiv  \hat P^J_{M0} \hat P^N\hat P^Z\vert \Phi(\mathbf{q})\rangle,
 \end{equation}
 with $\hat P^{J}_{MK}$,  $\hat{P}^{N, Z}$, and $\hat P^\pi$ are the projection operators that extract the component with the right angular momentum $J$, neutron number $N$, proton number $Z$, 
\begin{subequations}
\begin{align}
    \hat P^{J}_{MK}&=\dfrac{2J+1}{8\pi^2}\int d\Omega D^{J\ast}_{MK}(\Omega) \hat R(\Omega),\\
    \hat P^{N_\tau} &= \dfrac{1}{2\pi}\int^{2\pi}_0 d\varphi_{\tau}  e^{i\phi_{\tau}(\hat N_\tau-N_\tau)},
\end{align} 
\end{subequations} 
where  $N_\tau=N$ and $Z$ for neutrons and protons, respectively. The mean-field wave functions $\ket{\Phi(\mathbf{q})}$ are generated from the self-consistent relativistic mean-field plus Bardeen-Cooper-Schrieffer (RMF+BCS) theory ~\cite{Yao:2009PRC} with constraints on both the mass quadrupole moment  and pairing amplitude~\cite{Vaquero:2011PLB,Xiang:2020}
\beqn
\label{eq:energy_variation}
    \bra{\Phi}\hat{H}\ket{\Phi}
    &=&\bra{\Phi}\hat{H}_0\ket{\Phi}
    -\frac{1}{2}\lambda_Q\Bigg(\bra{\Phi}\hat Q_{20}\ket{\Phi}-q_{20}\Bigg)^2 \nonumber\\
    &&-\sum_{\tau=n,p} \lambda_{\tau}\Bigg(\bra{\Phi}\hat{N}\ket{\Phi}-N_\tau\Bigg)\nonumber\\
    &&-\xi_p\Bigg(\bra{\Phi}\hat P_{T=1}\ket{\Phi}-P_1\Bigg),
\eeqn
where $\bra{\Phi}\hat{H}_0\ket{\Phi}$ is given by the energy functional in the CDFT~\cite{Zhao:2010PRC,Meng:2016Book}, and $\lambda_Q, \lambda_\tau$ and $\xi_p$ are Lagrangian multipliers. The quadrupole moment operator is defined as $\hat Q_{20}  = r^2Y_{20}$, where $Y_{20}$ is a spherical harmonic function. The axial deformation parameter $\beta_2$ of the mean-field state $\ket{\Phi(\mathbf{q})}$ is determined by the expectation value of the quadrupole moment  $\beta_2= \frac{4\pi}{3AR^2} \bra{\Phi(\mathbf{q})}\hat Q_{20} \ket{\Phi(\mathbf{q})}$, where $R=1.2A^{1/3}$~fm with $A$ being the mass number. Following Refs.~\cite{Sieja:2004, Xiang:2020}, the last term in (\ref{eq:energy_variation}) is introduced to generate mean-field states with different isovector pairing amplitudes defined by the following operator 
\beqn
\hat P_{T=1}   = \frac{1}{2}\sum_{k>0}\Bigg(c^\dagger_{k}c^\dagger_{\bar k}+c_{\bar k}c_k\Bigg).
\eeqn
We find that the last constraint term on the pairing amplitude simply replaces the pairing gap $\Delta_{k}$ of $k$-th single-particle state in canonic basis, 
\begin{equation}
\Delta^\tau_{k} = \int \psi_{k}^{\dagger}(\boldsymbol{r}) \Delta_{\tau}(\boldsymbol{r}) \psi_{k}(\boldsymbol{r}) d \boldsymbol{r}
\end{equation}
 with $\Delta^\tau_{k}+\xi_p$ in the BCS equations~\cite{Ring:1980}, where the pairing field corresponding to  the density-independent $\delta$ force multiplied by a scaling factor $\chi$
 \beq
 \label{eq:pairing_force}
 V^{pp}_\tau(\bm{r}_1,\bm{r}_2)=\chi V^{pp}_\tau\delta(\bm{r}_1-\bm{r}_2)
 \eeq
 is given by
\begin{equation}
\Delta_{\tau}(\boldsymbol{r}) =\frac{V^{pp}_\tau}{2} \kappa_{\tau}(\boldsymbol{r})
\end{equation}
and the pairing tensor
\begin{equation}
\kappa_\tau(\boldsymbol{r})=-2 \sum^\tau_{k>0} f_{k} u_{k} \varv_{k}\left|\psi_{k}(\boldsymbol{r})\right|^{2}.
\end{equation}
 Thus, a continuous change of the parameter $\xi_p$ generates a set of BCS wave functions $\ket{\Phi(\mathbf{q})}$ labeled with different average pairing gaps 
\beqn
\label{eq:average_pairing_gap}
\Delta_{u\varv}=\frac{1}{2}(\Delta^n_{u\varv}+\Delta^p_{u\varv}),\quad 
\Delta^\tau_{u\varv}=\frac{\sum^\tau_{k>0} \Delta^\tau_k f_ku_k\varv_k}{\sum^\tau_{k>0} f_ku_k\varv_k},
\eeqn
where $\varv^2_k$ is the occupation probability of the $k$-th single-particle state, $u_k=(1-\varv^2_k)^{1/2}$, and $f_k$ is a cutoff function decreasing smoothly with the increase of single-particle energy~\cite{Bender:2003RMP,Yao:2009PRC}. 
Axial symmetry is imposed in the calculation. Thus, the obtained  mean-field wave functions $\ket{\Phi(\mathbf{q})}$ are labeled with two collective coordinates $(\beta_2, \Delta_{u\varv})$ for the intrinsic axial deformation parameter $\beta_2$ and pairing gap $\Delta_{u\varv}$, respectively. 
 
 The weight functions $f_\sigma^{J}(\mathbf{q})$ and the energies $E_\sigma^{J}$ of the states $\vert \Psi^{JNZ}_\sigma\rangle$ are the solutions of the Hill-Wheeler-Griffin (HWG) equation~\cite{Ring:1980}
\begin{equation}
\label{eq:HWG}
\sum_{\mathbf{q}'} \left[ \mathcal{H}^{J}_{00}(\mathbf{q},\mathbf{q}') - E_\sigma^{J}\mathcal{N}^{J}_{00}(\mathbf{q},\mathbf{q}')
          \right] \, f_\sigma^{J}(\mathbf{q}')
= 0,
\end{equation}
where the norm $\mathcal{N}^{J}_{00}$ and Hamiltonian $\mathcal{H}^{J}_{00}$ kernels are given by
\beqn
\mathcal{N}^{J}_{00}(\mathbf{q},\mathbf{q}') 
&=&\langle \Phi(\mathbf{q})\vert\hat P^J_{00} \hat P^N\hat P^Z\vert \Phi(\mathbf{q}')\rangle,\\
\mathcal{H}^{J}_{00}(\mathbf{q},\mathbf{q}')&=&\langle \Phi(\mathbf{q})\vert\hat H\hat P^J_{00} \hat P^N\hat P^Z\vert \Phi(\mathbf{q}')\rangle.
\eeqn
 In the energy kernel, the energy overlap is taking the same functional form as that of mean-field energy, but replacing the densities and currents with mixed/transition ones, where the bra and ket states are different~\cite{Yao:2010}.  

 The spectroscopic quadrupole moment $Q_{s}$ of the state $J^\pi_\sigma$ is defined as~\cite{Yao:2013PRC}
 \begin{eqnarray}
\label{eq:spectroscopic_quadrupole}
Q_{s}\left(J^\pi_{\sigma}\right) 
&=&\sqrt{\frac{16 \pi}{5}}\left\langle \Psi^{JM=JNZ}_\sigma \left|er^2 Y_{20}\right| \Psi^{JM=JNZ}_\sigma\right\rangle \nonumber\\
 &=&e\sqrt{\frac{16 \pi}{5}} \frac{\langle J J 20 \mid J J\rangle}{\sqrt{2 J+1}}\nonumber\\
 &&\times \sum_{\mathbf{q}\mathbf{q}'} f^{J\ast}_{\sigma}(\mathbf{q}^{\prime}) f^J_{\sigma}(\mathbf{q})\left\langle JNZ, \mathbf{q}'\left\|\hat Q_{2}\right\| JNZ, \mathbf{q}\right\rangle. 
 \end{eqnarray} 
 The electric quadrupole (E2) transition strength for $J^\pi_{\sigma_{i}} \rightarrow J^\pi_{\sigma_{f}}$ is determined by
  \begin{eqnarray}
\label{eq:BE2} 
&&B(E2; J^\pi_{\sigma_{i}} \rightarrow J^\pi_{\sigma_{f}})  \nonumber \\
&=&\frac{1}{2 J_{i}+1}\left|\sum_{\mathbf{q}^{\prime}, \mathbf{q}} f_{\sigma_{f}}^{J_{f}^{*}}\left(\mathbf{q}^{\prime}\right)\left\langle J_{f} NZ, \mathbf{q}^{\prime}\left\|\hat{Q}_{2}\right\| J_{i} NZ, \mathbf{q}\right\rangle f_{\sigma_{i}}^{J_{i}}(\mathbf{q})\right|^{2},
 \end{eqnarray} 
where the reduced matrix element in (\ref{eq:spectroscopic_quadrupole}) and (\ref{eq:BE2}) contributed solely from protons is determined as 
  \begin{eqnarray}
 \left\langle J_{f}NZ, \mathbf{q}^\prime|| \hat{Q}_{2}|| J_{i}NZ, \mathbf{q}\right\rangle  
&=& \left(2 J_{f}+1\right)  (-1)^{J_{f}}\sum^2_{\mu=-2}\left(\begin{array}{ccc}
J_{f} & 2 & J_{i} \\
0 & \mu & -\mu
\end{array}\right)\nonumber\\
&&\times \bra{\Phi(\mathbf{q}^{\prime})} r^2Y_{2\mu}  \hat P^{J_i}_{-\mu0} \hat{P}^{N} \hat{P}^{Z}\ket{\Phi(\mathbf{q})}. \nonumber\\
 \end{eqnarray}

\subsection{Nuclear matrix elements of $0\nu\beta\beta$  decay}

 In the mechanism of exchanging either light ($\alpha=\nu$) or heavy ($\alpha=N$) Majorana neutrinos, the half-life $T^{0\nu}_{1/2}$ for the $0^+_1$ to $0^+_1$ transition can be factorized as below~\cite{Simkovic:1999,Hyvarinen:2015,Song:2017}
\begin{equation} 
 [T^{0\nu}_{1/2}]^{-1}=G_{0\nu}g^4_A\eta_{\alpha}^2 |M^{0\nu}_{\alpha}|^2,
\end{equation}
where $g_A\simeq 1.26$ is the axial vector coupling constant, and the phase space factor $G_{0\nu}$ can be determined rather precisely~\cite{Kotila:2012}. The quantity $\eta_{\alpha}$ describes the physics beyond the standard model~\cite{Simkovic:1999,Hyvarinen:2015}. 
\begin{itemize}
    \item For the mechanism of exchanging light neutrinos, the $\eta_{\nu}$ factor is related to the masses of light neutrinos
\begin{equation}
    \eta_{\nu}=\bigg|\frac{\braket{m_{\beta\beta}}}{m_e}\bigg|=\bigg|\frac{\sum_{\nu_j=1}^3U_{e\nu_j}^2m_{\nu_j}}{m_e}\bigg|
\end{equation}
\item For the mechanism of exchanging heavy neutrinos, the $\eta_{N}$ factor is related to the masses of heavy neutrinos
\begin{equation}
\eta_{N}=\bigg|\frac{m_p}{\langle M_{\beta\beta}\rangle}\bigg|=\bigg|\sum_{N_j=1}^3\frac{U_{eN_j}^{2}m_p}{M_{N_j}}\bigg|.
\end{equation}
\end{itemize}
In the above expression, $m_e=0.511$ MeV ($m_p=0.938$ GeV) is the electron (proton) mass. The  $U_{e\nu}$ and $U_{eN}$ are the elements of the neutrino mixing matrix that connect the electron flavor eigenstate to the mass eigenstates of light and heavy neutrinos, respectively. The effective neutrino masses of light and heavy Majorana neutrinos are defined as
\begin{equation}
\braket{m_{\beta\beta}}=\sum_{\nu_j=1}^3U_{e\nu_j}^2m_{\nu_j},\quad 
\braket{M^{-1}_{\beta\beta}}= \sum_{N_j=1}^3\frac{U_{eN_j}^{2}}{M_{N_j}}.
\end{equation}
 
The NME is computed with the wave functions for the initial and final nuclei,
\begin{equation} 
  M^{0\nu}_{\alpha} = \bra{\Psi_F}\hat{\mathcal O}^{0\nu}_{\alpha}\ket{\Psi_I}.
\end{equation}
The $0\nu\beta\beta$-decay operator is derived from the second-order weak Hamiltonian with charge-exchange nucleonic and leptonic currents. In the {\em closure approximation}, the transition operator can be written as follows~\cite{Yao:2021Review}
\begin{eqnarray}
\label{eq:operator}
	\hat{\mathcal O}^{0\nu}_{\alpha}&=&\frac{4\pi R}{g_A^2} \iint d^3 x_1 d^3 x_2 \int\frac{d^3q}{(2\pi)^3}\, h_\alpha(q)\nonumber\\
	&\times & \hat{\mathcal{J}}_\mu^\dagger(\bm x_1)\hat{\mathcal{J}}^{\mu\dagger}(\bm x_2) e^{i\bm q\cdot(\bm x_1-\bm x_2)},
\end{eqnarray}
with $R=1.2A^{1/3}$ fm. The neutrino potential $h_\nu(q)$ of exchanging light neutrinos  is
\begin{eqnarray}
\label{eq:hofq-light}
	h_\nu(q)&=&[q(q+E_d)]^{-1}\,,\\
	E_d&\equiv & \bar E-(E_I+E_F)/2\,,\notag
\end{eqnarray}
where  $E_{I(F)}$ corresponds to the energy of initial (final) nuclear state, and $\bar E$ is the average energy of intermediate states. The value of $E_d$ is chosen according to the empirical formula $E_d=1.12A^{1/2}$ MeV~\cite{Haxton1984PPNP}. The neutrino potential $h_N(q)$ of exchanging heavy neutrinos is
\begin{eqnarray}
\label{eq:hofq-heavy}
	h_N(q)=(m_pm_e)^{-1}.
\end{eqnarray}

Within the impulse approximation, the one-body charge-changing nucleon current operator can be written into sencond quantization form
\beq
\hat{\mathcal{J}}_\mu^\dagger(0)
=\sum_{pp'} \bra{N(p')} \mathcal{J}_\mu^\dagger(0) \ket{N(p)}a^\dagger_{p'}a_p,
\eeq
where $p, p'$ are the momenta of nucleons in free space. The matrix element reads
\begin{eqnarray}
\label{eq:current_matrix}
\bra{N(p')}\mathcal{J}_\mu^\dagger(0)
\ket{N(p)}
\equiv\bar\psi(p') \Gamma_\mu(q)\tau_-\psi(p)
\end{eqnarray} 
Here $\tau_-$ is the $2\times2$ matrix representation of the isospin lowering operator, changing neutron to proton, and $\psi(p)$ is composed of two Dirac spinors for neutron and proton wave functions. $q=p-p'$ is the transferred momentum. The coupling vertex reads~\cite{Simkovic99,Yao:2021Review},
\begin{equation}
\label{eq:gammamu}
\begin{aligned}
    \Gamma_\mu(q)&=g_V(q^2)\gamma_\mu-i g_M(q^2)\frac{\sigma_{\mu\nu}}{2m_p}q^\nu \\
    &-g_A(q^2)\gamma_\mu \gamma_5+g_P (q^2)q_\mu \gamma_5,
\end{aligned}
\end{equation} 
where  $\sigma_{\mu\nu}=\frac{i}{2}\left[\gamma_\mu,\gamma_\nu\right]$ and $g_i(q^2)$ are form factors~\cite{Simkovic:1999,Song:2014,Yao:2021Review}.

Substituting the above expression into (\ref{eq:operator}), one finds that the NME is composed of five terms: vector coupling (VV), axial-vector coupling (AA), interference of the axial-vector and induced pseudoscalar coupling (AP), the induced pseudoscalar coupling (PP), and weak-magnetism coupling (MM)  terms, which are related to the products of two current operators $\hat{\mathcal{J}}_\mu^\dagger\hat{\mathcal{J}}^{\mu\dagger}$ with the following forms~\cite{Song:2014,Yao:2015},
\begin{subequations}
\label{twocurrentR}
\begin{eqnarray}
\label{VV}
VV: &&g_V^2(q^2)\left(\bar\psi\gamma_\mu\tau_-\psi\right)^{(1)}\left(\bar\psi\gamma^\mu\tau_-\psi\right)^{(2)},\\
\label{AA}
AA: &&g_A^2(q^2)\left(\bar\psi\gamma_\mu\gamma_5\tau_-\psi\right)^{(1)}\left(\bar\psi\gamma^\mu\gamma_5\tau_-\psi\right)^{(2)},\\
\label{AP}
AP: &&2g_A(q^2)g_P(q^2)\left(\bar\psi\bm \gamma\gamma_5\tau_-\psi\right)^{(1)}\left(\bar\psi \bm q\gamma_5\tau_-\psi\right)^{(2)},\\
\label{PP}
PP: &&g_P^2(q^2)\left(\bar\psi \bm q\gamma_5\tau_-\psi\right)^{(1)}\left(\bar\psi \bm q\gamma_5\tau_-\psi\right)^{(2)},\\
\label{MM}
MM: &&g_M^2(q^2)\left(\bar\psi\frac{\sigma_{\mu i}}{2m_p}q^i\tau_-\psi\right)^{(1)}\left(\bar\psi\frac{\sigma^{\mu j}}{2m_p}q_j\tau_-\psi\right)^{(2)}.
\end{eqnarray}
\end{subequations}

With the nuclear wave functions constructed in Eq.~(\ref{eq:gcmwf}), the total NME can be written as
\begin{eqnarray}
\label{eq:NME}
M^{0\nu}_\alpha
&=&\sum_{\mathbf{q}_I,\mathbf{q}_F} f^{0^+_F}_1(\mathbf{q}_F)f^{0^+_I}_1(\mathbf{q}_I) 
\sqrt{\mathcal{N}^{J=0}_{00}(\mathbf{q}_I,\mathbf{q}_I) \mathcal{N}^{J=0}_{00}(\mathbf{q}_F,\mathbf{q}_F) }\nonumber\\
&&\times\tilde{M}^{0\nu}_\alpha(\mathbf{q}_F, \mathbf{q}_I),
\end{eqnarray}
with the normalized NME defined as
\begin{equation}
\label{eq:normalizedNME}
\tilde{M}^{0\nu}_\alpha(\mathbf{q}_F, \mathbf{q}_I) 
= \frac{\bra{\Phi_F(\mathbf{q}_F)}{\mathcal O}^{0\nu}_\alpha\hat P^{J=0} \hat P^{N_I}\hat P^{Z_I}\ket{\Phi_I(\mathbf{q}_I)}}{\sqrt{\mathcal{N}^{J=0}_{00}(\mathbf{q}_I,\mathbf{q}_I) \mathcal{N}^{J=0}_{00}(\mathbf{q}_F,\mathbf{q}_F) }},
\end{equation} 
where $\ket{\Phi_{I/F}(\mathbf{q})}$ are the mean-field wave functions of  initial and final nuclei with collective parameters $\mathbf{q}_I$ and $\mathbf{q}_F$, respectively. The nucleon wave function $\psi$ in the matrix element (\ref{eq:current_matrix}) of current operator corresponds to a single-particle state of neutron or proton inside atomic nucleus.

 The short-range correlation (SRC) effect between nucleons on the NME of $0\nu\beta\beta$ decay  is taken into account  by multiplying a Jastrow correlation function,
\begin{eqnarray}
\label{eq:Fofr}
	F(r)=1-ce^{-ar^2}(1-br^2),
\end{eqnarray}
onto the transition operator~\cite{Jastrow:1955PR,Miller:1976AP}
 \beq
 \hat{\mathcal O}^{0\nu}(r)\to F(r)\hat{\mathcal O}^{0\nu}(r)F(r).
 \eeq
where $r\equiv |\mathbf{x}_1-\mathbf{x}_2|$ is the distance of two nucleons.  The CD-Bonn parameters $a=1.52$ fm$^{-2}$, $b=1.88$ fm$^{-2}$, and $c=0.46$ are employed. See Refs.~\cite{Simkovic:2009PRC,Song:2017} for details.

\section{Results and discussions}
\label{sec:results}
 
In the mean-field calculations, parity, $x$-simplex symmetry, and time-reversal invariance are imposed. The Dirac equations for neutrons and protons are solved by expanding the large and small components of Dirac spinor in the basis of eigenfunctions of a three-dimensional harmonic oscillator (H.O.) in Cartesian coordinate with $n_f$ major shells. Table ~\ref{tab:NoS} shows the NMEs $M^{0\nu}_{\nu/N}$ of $0\nu\beta\beta$ decay from the MR-CDFT calculation with only quadrupole shape fluctuation and  $n_f=6, 8, 10, 12$.  One can see that the NME varies only 0.6\% when the $n_f$ increases from 10 to 12. Therefore,  $n_f=10$ is adopted in the subsequent calculations. The relativistic  point-coupling density functional PC-PK1~\cite{Zhao:2010PRC} is adopted. We note that in the previous studies with MR-CDFT~\cite{Song:2014,Yao:2015,Song:2017,Yao:2016PRC}, the pairing strength parameters $V^{pp}_\tau$ were chosen as $-314.550$ MeV fm$^3$ and $-346.500$ MeV fm$^3$ for neutrons and protons, respectively, which were determined by fitting to the neutron and proton average pairing gaps as functions of deformation parameter $\beta_2$ in $^{150}$Nd, and $^{150}$Sm provided by the separable finite-range pairing force~\cite{Tian:2009PLB}.  In this work, these pairing strengths will be adjusted to the excitation energy of $2^+_1$ state of each nucleus by multiplying a scaling factor $\chi$ for both neutrons and protons, c.f. (\ref{eq:pairing_force}). As demonstrated in this work, the excitation energy of $2^+_1$ state is linearly correlated to the scaling factor of the pairing strengths. Thus, it is justified to determine the pairing strengths using the data of excitation energies, instead of the data of odd-even mass difference. The Gauss-Legendre quadrature is used for integrals over the Euler angle $\theta$ and gauge angle $\varphi_{\tau=n,p}$ in the calculations of the quantum-number projected norm and Hamiltonian kernels. To accelerate the MR-CDFT calculation with both shape fluctuation and isovector pairing fluctuation, the recently proposed orthogonality condition method~\cite{Romero:2021PRC,Zhang:2023PRC} is employed to select the optimal configurations relevant for nuclear low-lying states and the NMEs of $0\nu\beta\beta$ decay.

\begin{table}[]
	\centering  
        \tabcolsep=11pt 
	\caption{Convergence of nuclear matrix elements for the $0\nu\beta\beta$ decay of $\nuclide[130]{Te}$ in the mechanism of exchanging either light ($M^{0\nu}_\nu$) or heavy ($M^{0\nu}_N$) Majorana neutrinos with respect to the number of H.O. major shells in the expansion of single-particle wave functions. }  
	\label{tab:NoS}  
	\begin{tabular}{c|cccc}  
		\toprule
		  $n_f$ & 6 & 8 & 10 &12 \\ 
		\hline
            $M^{0\nu}_{\nu}$ & 5.057& 5.214& 4.893& 4.916 \\ 
            $M^{0\nu}_{N}$ &253.844& 273.553& 257.644& 259.389 \\
            \bottomrule
	\end{tabular}
\end{table}

\subsection{Calibration of pairing strengths and NMEs of $0\nu\beta\beta$ decay with excitation energies }

\begin{figure}[bt]
     \centering \includegraphics[width=\columnwidth]{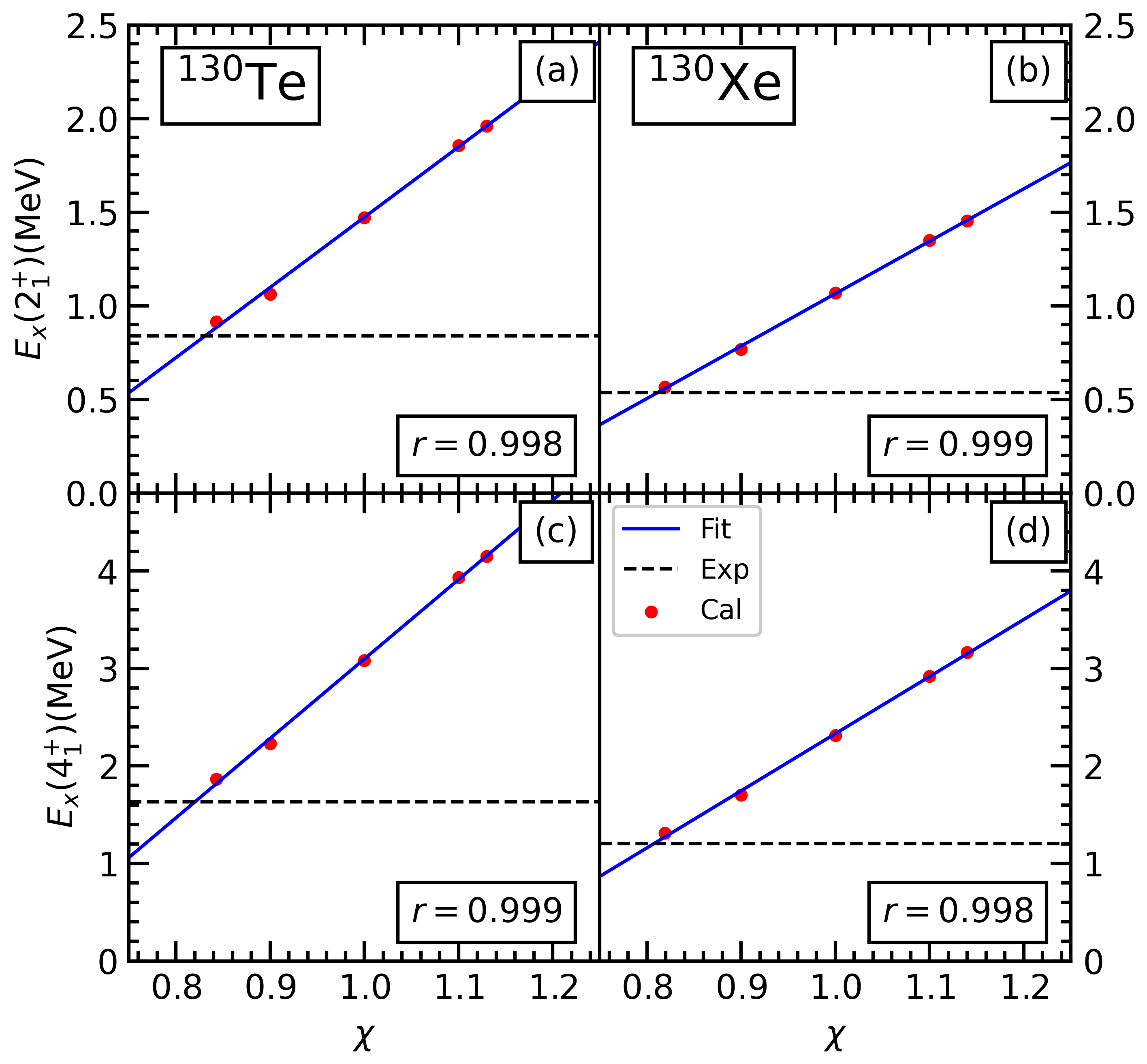} 
     \caption{The excitation energies of $2^+_1$, $4^+_1$ states as a function of the scaling factor of the pairing strengths for $^{130}$Te and $^{130}$Xe. The Pearson correlation coefficient  $r$ is provided for each case. The dashed lines mark the location of corresponding data taken from Ref.~\cite{NNDC}. See main text for details.}
     \label{fig:Ex4TeXe_change_pairing}
\end{figure}

\begin{figure}
     \centering \includegraphics[width=\columnwidth]{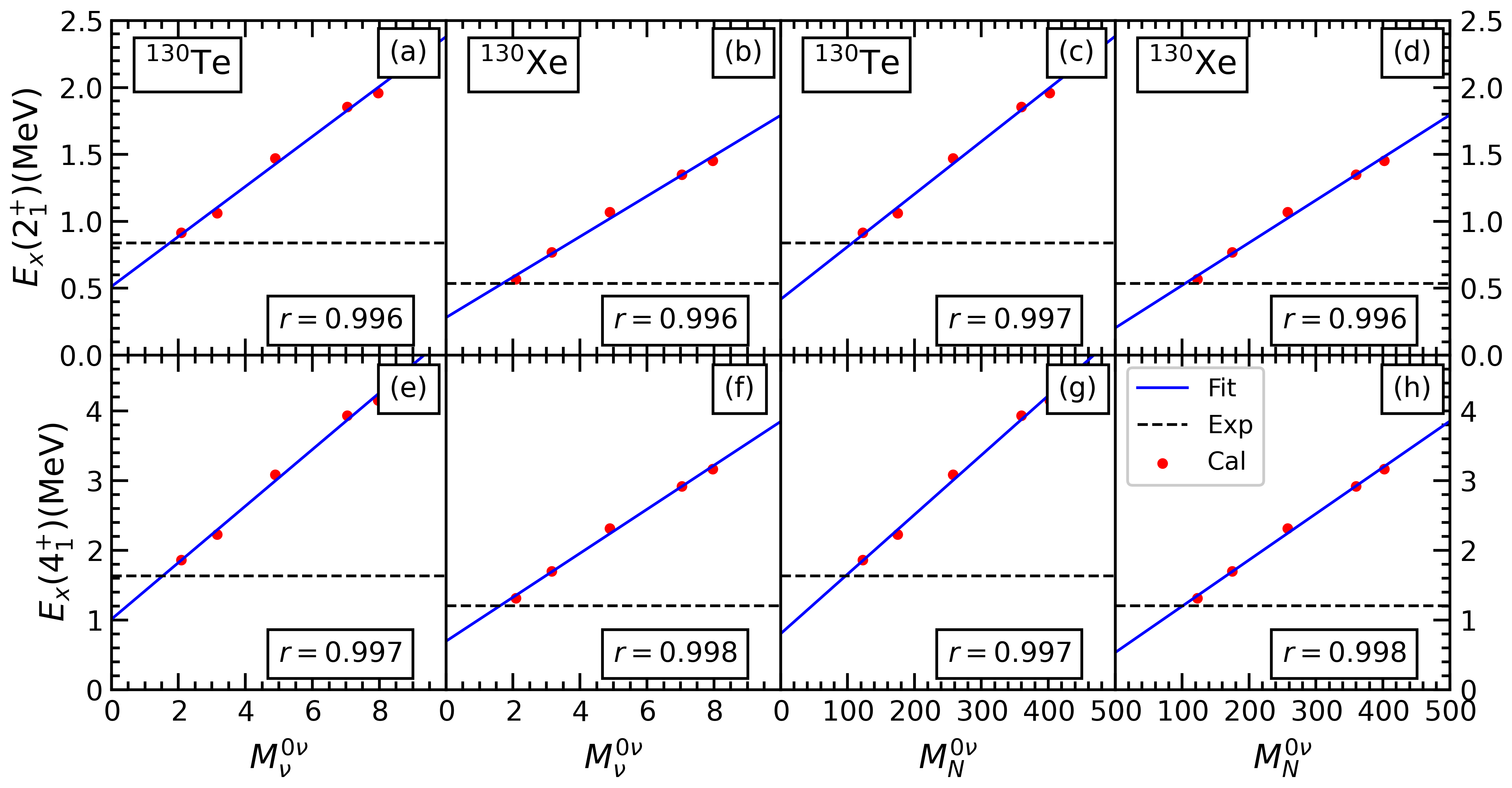} 
     \caption{Correlation between excitation energies of $2^+_1$, $4^+_1$ states and the NMEs $M^{0\nu}_{\nu/N}$ in $^{130}$Te and $^{130}$Xe. The Pearson correlation coefficient  $r$ is provided. The dashed lines indicate the location of the corresponding data from Ref.~\cite{NNDC}.}
     \label{fig:correlation_Ex_NME_TeXe}
\end{figure} 

We first examine the correlation between the excitation energies of nuclear low-lying states and nucleon isovector pairing strengths in the MR-CDFT. To this end, we carry out GCM calculations with the mixing of different axially-deformed configurations using different values of the scaling factor $\chi$ in the isovector pairing strengths $V^{pp}_{\tau}$. For the illustration purpose, we only show the excitation energies of $2^+_1$ and $4^+_1$ states for \nuclide[130]{Te} and \nuclide[130]{Xe} in Fig.~\ref{fig:Ex4TeXe_change_pairing} as a function of $\chi$.  The value of the Pearson's correlation coefficient $r$ defined as 
\beq
r=\frac{\sum_{i=1}^N\left(x_i-\bar{x}\right)\left(y_i-\bar{y}\right)}{\sqrt{\sum_{i=1}^N\left(x_i-\bar{x}\right)^2 \sum_{i=1}^N\left(y_i-\bar{y}\right)^2}}
\eeq
 is also given to demonstrate the linear correlation between the two variables $(x, y)$. One can see that the correlation coefficient $r$ is very close to one. It indicates that the excitation energies of $2^+_1$ and $4^+_1$ states are linearly correlated with the pairing strengths in both nuclei. A similar linear correlation is also observed in other four candidate nuclei.  In the mean time, we find that the excitation energies are also linearly correlated with the predicted NMEs of $0\nu\beta\beta$ decay, as shown in Fig.~\ref{fig:correlation_Ex_NME_TeXe}. In other words, the NMEs of $0\nu\beta\beta$ decay are sensitive to the scaling factor $\chi$, i.e., the isovector pairing strengths. A similar correlation was also found in the recent ISM study based on a pairing-plus-quadruple Hamiltonian~\cite{Yoshinaga:2018,Higashiyama:2020}. These findings provide a strong foundation to calibrate the pairing strengths and finally the NMEs of $0\nu\beta\beta$ decay in MR-CDFT calculations using the data of excitation energies. Table~\ref{tab:pairing_strength} presents the scaling factors for the ten nuclei of interest, which are determined by fitting to the data of the excitation energies of $2^+_1$ states~\cite{NNDC}.

\begin{table}[]
    \centering
    \renewcommand{\arraystretch}{1.0}
    \tabcolsep=2pt 
    \caption{The employed values of the scaling factor $\chi$ (\ref{eq:pairing_force}) multiplied to the pairing strengths for the nuclei of interest.}
    \begin{tabular}{cccccccccc}
      \toprule
       $^{76}$Ge& $^{76}$Se& $^{82}$Se& $^{82}$Kr& $^{100}$Mo& $^{100}$Ru& $^{130}$Te  & $^{130}$Xe& $^{136}$Xe& $^{136}$Ba \\
       \hline
         $0.671$& $0.852$& $0.836$& $0.736$& $0.950$& $0.985$& $0.831$& $0.811$& $0.630$& $0.807$\\
       \bottomrule
    \end{tabular}
    \label{tab:pairing_strength}
\end{table}

\begin{figure}
     \centering \includegraphics[width=\columnwidth]{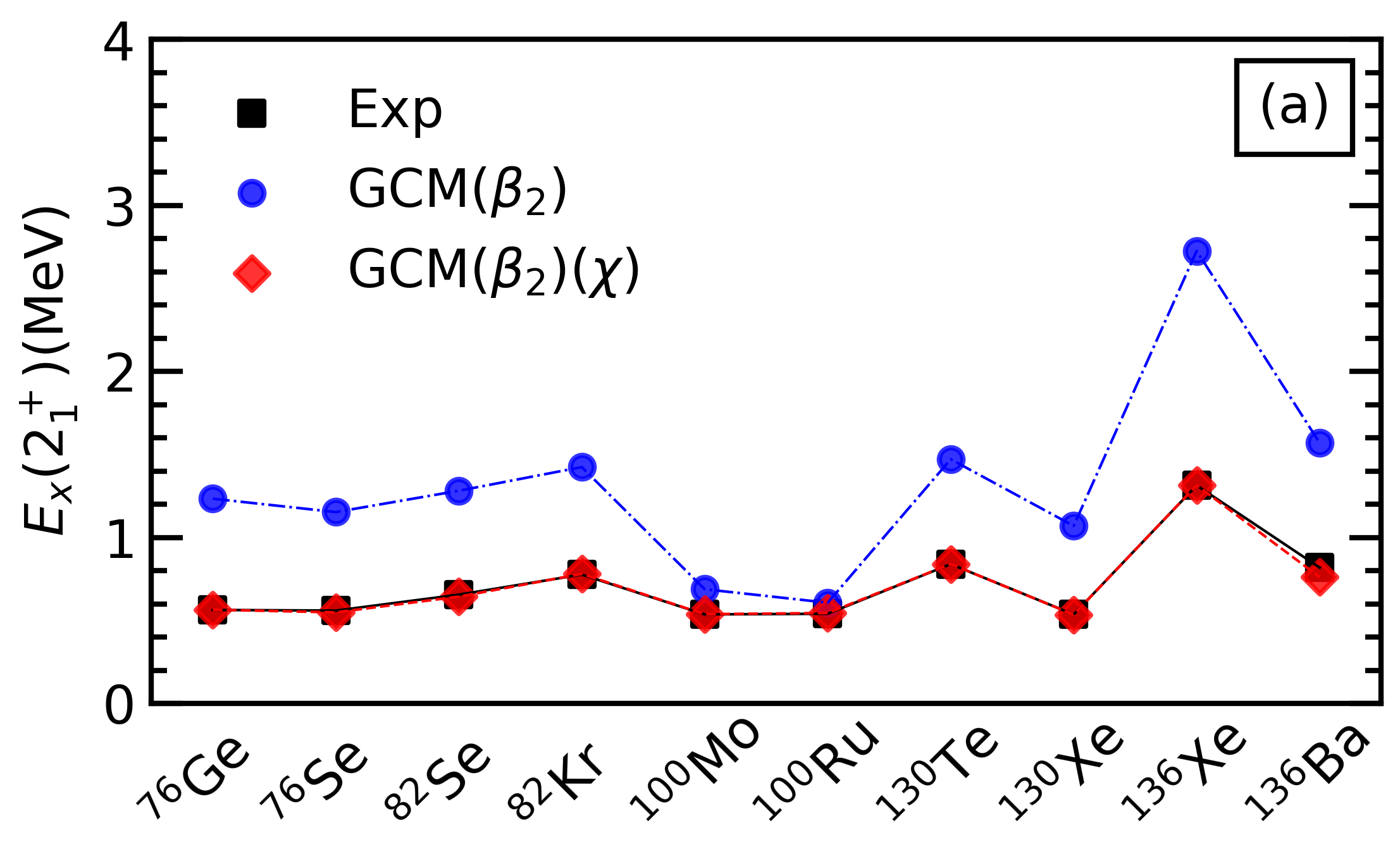} 
     \centering \includegraphics[width=\columnwidth]{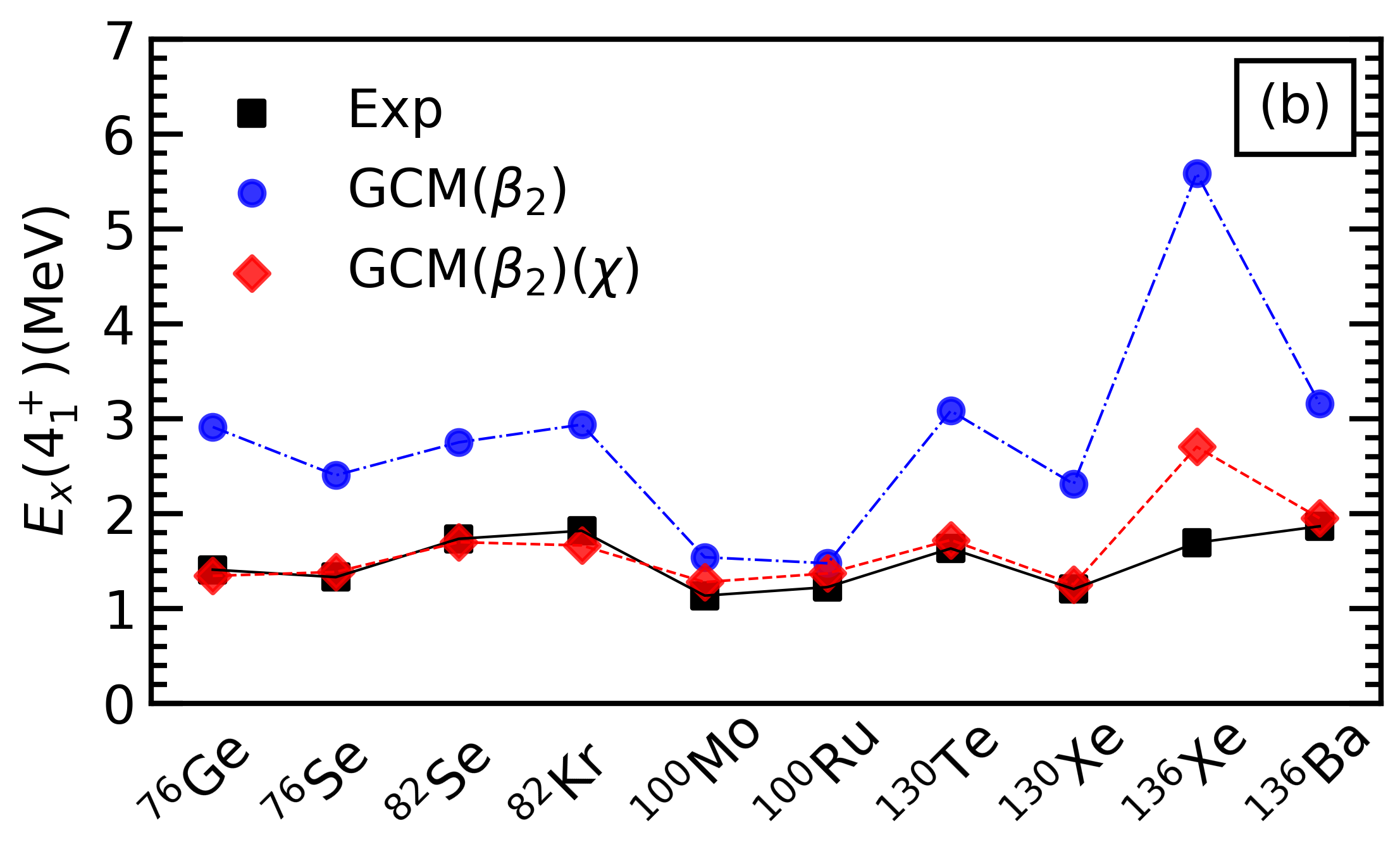} 
     \caption{Excitation energies of (a)  $2^+_1$ states and (b) $4^+_1$ states in the candidate nuclei of $0\nu\beta\beta$ decay from the GCM calculation with and without multiplying the pairing strengths by the scaling factor $\chi$, in comparison with corresponding data~\cite{NNDC}.}
     \label{fig:excitation_energies}
 \end{figure} 

 \begin{figure}
     \centering \includegraphics[width=\columnwidth]{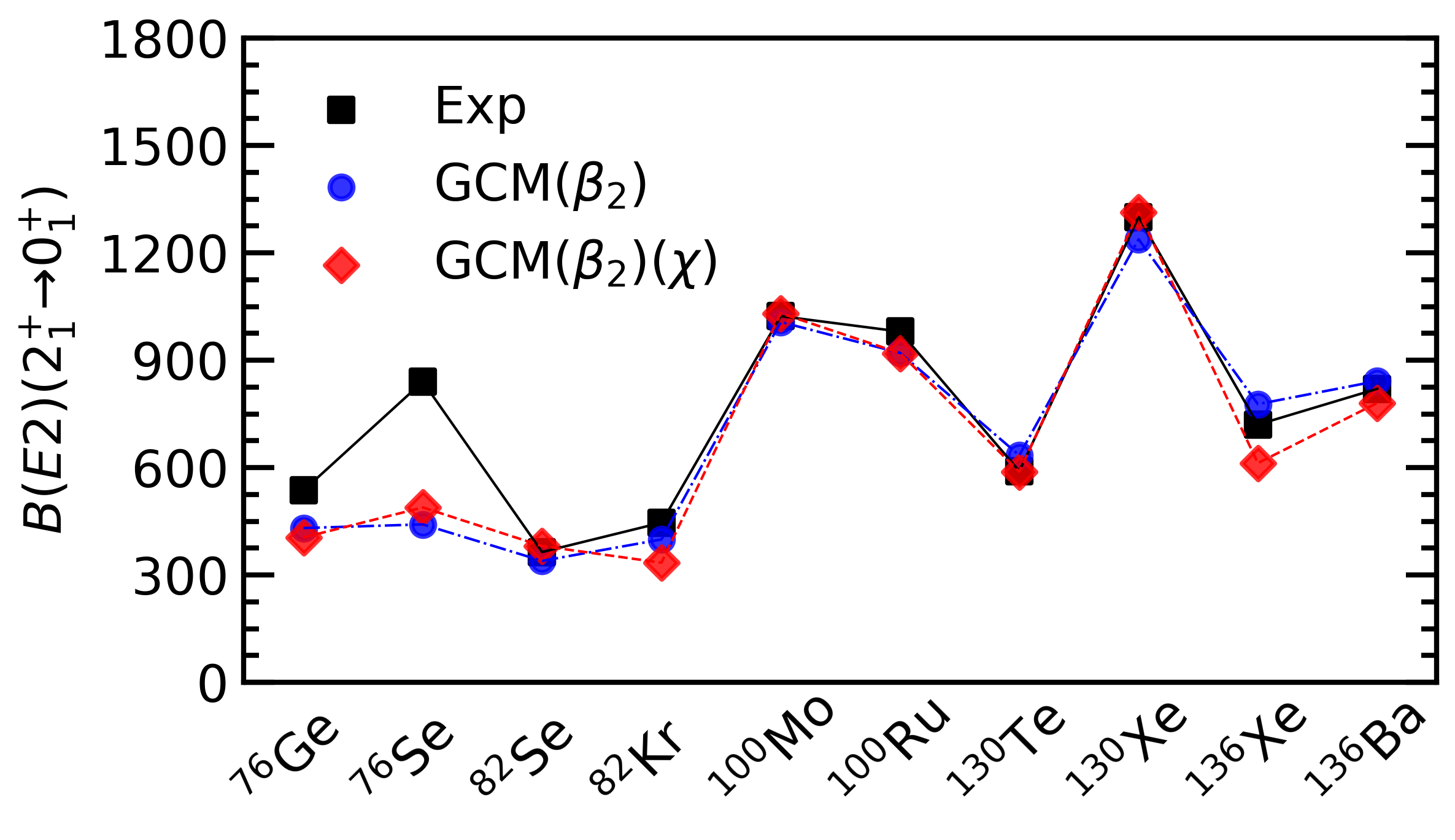} 
     \caption{Same as Fig.~\ref{fig:excitation_energies}, but for the electric quadrupole transition strength $B(E2:2^+_1\to 0^+_1)$ from $2^+_1$ state to $0^+_1$ state.}
     \label{fig:BE2}
 \end{figure}

 \begin{table*}[bt]
    \centering
    \renewcommand{\arraystretch}{1.0}
    \tabcolsep=4pt 
    \caption{The NMEs $M^{0\nu}_{\nu/N}$ of $0\nu\beta\beta$ decay and the limits of the effective neutrino mass $|\braket{m_{\beta\beta}}|$ (meV) and $|\braket{M_{\beta\beta}}|^{-1}$ ($\times 10^{6}$ GeV) for five candidate nuclei with and without multiplying the pairing strengths by the scaling factor $\chi$. The transition operator corresponding to the mechanism of exchanging either light or heavy Majorana neutrino is employed. The lower half-life limits for five candidate nuclei are taken from the latest experimental measurements~\cite{GERDA:2020,Azzolini:2022PRL,Augier:2022EPJC,Adams:2022fgm,Abe:2023PRL}. }
    \begin{tabular}{cc|ccccccc|ccccccc}
      \toprule
       \multirow{2}{*}{Isotopes} &\multirow{2}{*}{ Models} & \multicolumn{6}{c}{$M^{0\nu}_{\nu}$(light neutrino)} && \multicolumn{6}{c}{$M^{0\nu}_{N}$(heavy neutrino)}&\\
       \cline{3-16}
        & & VV & AA & AP &PP &MM & Tot&$|\braket{m_{\beta\beta}}|$& VV & AA  & AP&PP&MM & Tot &  $|\braket{M_{\beta\beta}}|^{-1}$ \\
        \hline
       \multirow{2}{*}{\nuclide[76]{Ge}}&GCM($\beta_2$) &$1.30$& $5.83$& $-1.85$& $0.73$& $0.22$& $6.23$ &$<113.0$&$64.45$& $295.28$& $-186.48$& $103.25$& $14.98$&$291.49$ &$>284.2$\\
       &GCM($\beta_2$)($\chi$) & $0.48$& $2.22$& $-0.70$& $0.28$& $0.09$&  $2.37$ &$<297.1$& $25.81$ & $119.92$ & $-73.22$& $40.89$ &$6.62$& $120.02$ &$>117.0$\\ 
        \hline
       \multirow{2}{*}{\nuclide[82]{Se}}&GCM($\beta_2$) &$1.17$& $5.09$& $-1.68$& $0.66$& $0.20$& $5.43$&$<276.7$&$59.03$& $270.22$& $-174.05$& $93.50$& $13.71$&$262.42$ &$>83.8$\\
       &GCM($\beta_2$)($\chi$) &$0.48$& $2.36$& $-0.78$& $0.30$& $0.10$&  $2.46$ &$<610.8$&$28.91$& $137.71$& $-83.24$& $43.86$& $7.34$& $134.58$ &$>43.0$\\ 
       \hline
       \multirow{2}{*}{\nuclide[100]{Mo}}&GCM($\beta_2$) &$1.26$& $6.47$& $-2.01$& $0.83$& $0.24$& $6.78$ &$<283.1$&$72.55$& $330.16$& $-215.71$& $121.35$& $18.29$&$326.63$ &$>82.3$\\
       &GCM($\beta_2$)($\chi$) &$1.10$& $5.73$& $-1.77$& $0.73$& $0.21$&  $6.01$ &$<319.4$& $64.13$& $292.30$& $-190.44$& $106.91$& $16.27$& $289.17$ &$>72.9$\\ 
       \hline
       \multirow{2}{*}{\nuclide[130]{Te}}&GCM($\beta_2$) &$0.93$ &$4.66$ &$-1.51$&$0.62$&$0.19$& $4.89$ &$<118.7$&$53.88$ &$255.56$&$-157.04$&$89.40$&$15.85$ &$257.64$ &$>213.2$\\
       &GCM($\beta_2$)($\chi$) & $0.32$&$1.83$&$-0.59$&$0.25$&$0.08$&  $1.89$ &$<307.1$&$22.20$&$109.57$& $-62.00$&$35.86$&$7.42$&  $113.06$ &$>93.6$\\ 
       \hline
       \multirow{2}{*}{\nuclide[136]{Xe}}&GCM($\beta_2$) &$0.81$& $4.08$& $-1.31$& $0.53$& $0.16$& $4.27$  &$<41.6$&$46.75$& $223.31$& $-135.93$& $75.68$& $13.60$&$223.41$ &$>586.2$\\
       &GCM($\beta_2$)($\chi$) &$0.37$& $2.20$& $-0.66$& $0.26$& $0.10$&  $2.27$ &$<78.2$&$24.61$& $124.83$& $-66.89$& $37.26$& $8.14$& $127.95$ &$>335.7$\\ 
       \bottomrule
    \end{tabular}
    \label{tab:pairingf-NME}
\end{table*}

 Figure~\ref{fig:excitation_energies} displays the excitation energies of $2^+_1$ and $4^+_1$ states in the five pairs of $0\nu\beta\beta$-decay candidate nuclei from the MR-CDFT calculation with the pairing strengths of the previous studies~\cite{Song:2014}, labeled as GCM($\beta_2$), and those multiplied by the scaling factors, labeled as GCM($\beta_2$)($\chi$). As shown in the previous study~\cite{Yao:2015} and Fig.~\ref{fig:excitation_energies}, the excitation energies of $2^+_1$ states and $4^+_1$ states are systematically overestimated in the GCM($\beta_2$) calculation by a factor ranging from 1.3 to 2.2. After introducing the scaling factor parameter $\chi$ in the GCM($\beta_2$)($\chi$) calculation, we are also able to reproduce the excitation energies of the $4^+_1$ states excellently.  Previous studies have demonstrated that part of the overestimation of the excitation energies can be reduced by including cranked states~\cite{Sabbey:2007PRC,Rodriguez:2015PRC,Bally:2014PRL,Borrajo:2015}, because cranked configurations take into account alignment effects which increase the angular momentum without changing pairing correlations. If this effect is included in the GCM calculation, one anticipates the scaling factors $\chi$ of the isovector pairing strengths would be slightly larger. Its impact on the predicted NME will be discussed later. Of particular interest is that the description of the $E2$ transition strengths $B(E2;2^+_1\to 0^+_1)$ remains roughly similar with the presence of the scaling factor, as shown in Fig.~\ref{fig:BE2}.
 
 Table~\ref{tab:pairingf-NME} lists the decomposition of the $0\nu\beta\beta$-decay NMEs $M^{0\nu}_{\alpha}$ from the GCM($\beta_2$) and GCM($\beta_2$)$(\chi)$ calculations for the five candidate nuclei.  We note that the observed slight difference between the previous NMEs~\cite{Song:2017} and the value by the GCM($\beta_2$) in Tab.~\ref{tab:pairingf-NME} is from the different choices of parametrizations of the short-range correlation. Here, we choose the CD-Bonn parametrization, instead of the Argonne V18~\cite{Simkovic:2009PRC}. In Table~\ref{tab:pairingf-NME}, it is evidently seen that the NMEs of exchanging both light and heavy neutrinos are significantly reduced in the calculation with the scaling factors. Quantitatively, the total NME is reduced by a factor ranging from $12\%$ to $62\%$. Using the current half-life limit of each candidate nucleus, we find that the reduction of the NME increases the upper  (lower) limit on the effective masses of light (heavy) neutrinos by $13\%$-$163\%$.

\subsection{Impact of nucleon isovector pairing fluctuation}

\begin{figure}
	\centering
	\includegraphics[width=\columnwidth]{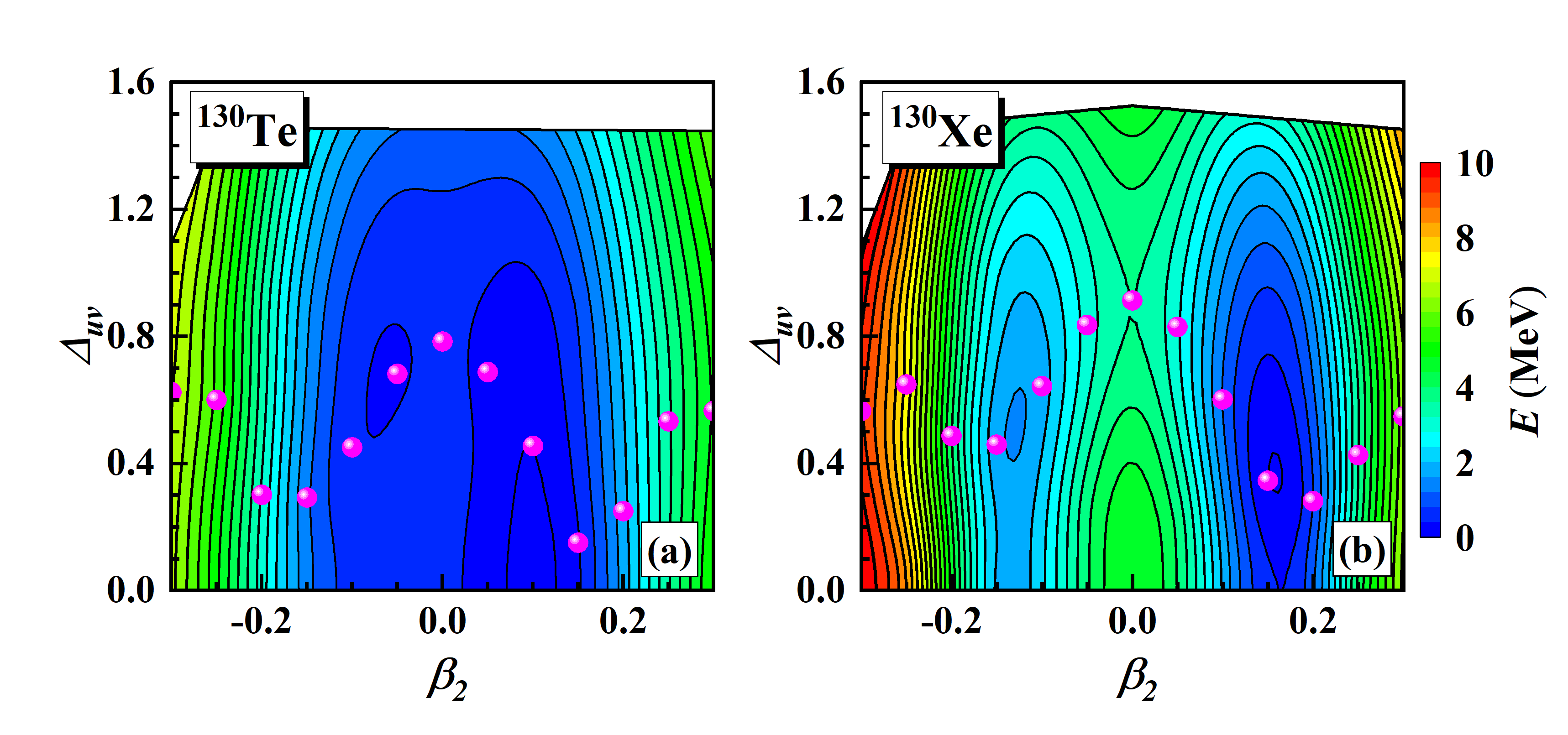} 
	\caption{The energies of mean-field states (normalized to the global energy minimum) from the CDFT calculation with the adjusted pairing strengths for (a) $^{130}$Te  and (b) $^{130}$Xe  as a function of quadrupole deformation parameter $\beta_2$ and average pairing gap $\Delta_{u\varv}$ of protons and neutrons. The red dots mark the location of $\Delta_{u\varv}$ in the states from the mean-field calculation with a constraint only on the quadrupole deformation parameter $\beta_2$.
	}
	\label{fig:MF_PES_TeXe}
\end{figure}

\begin{figure}[]
     \centering \includegraphics[width=\columnwidth]{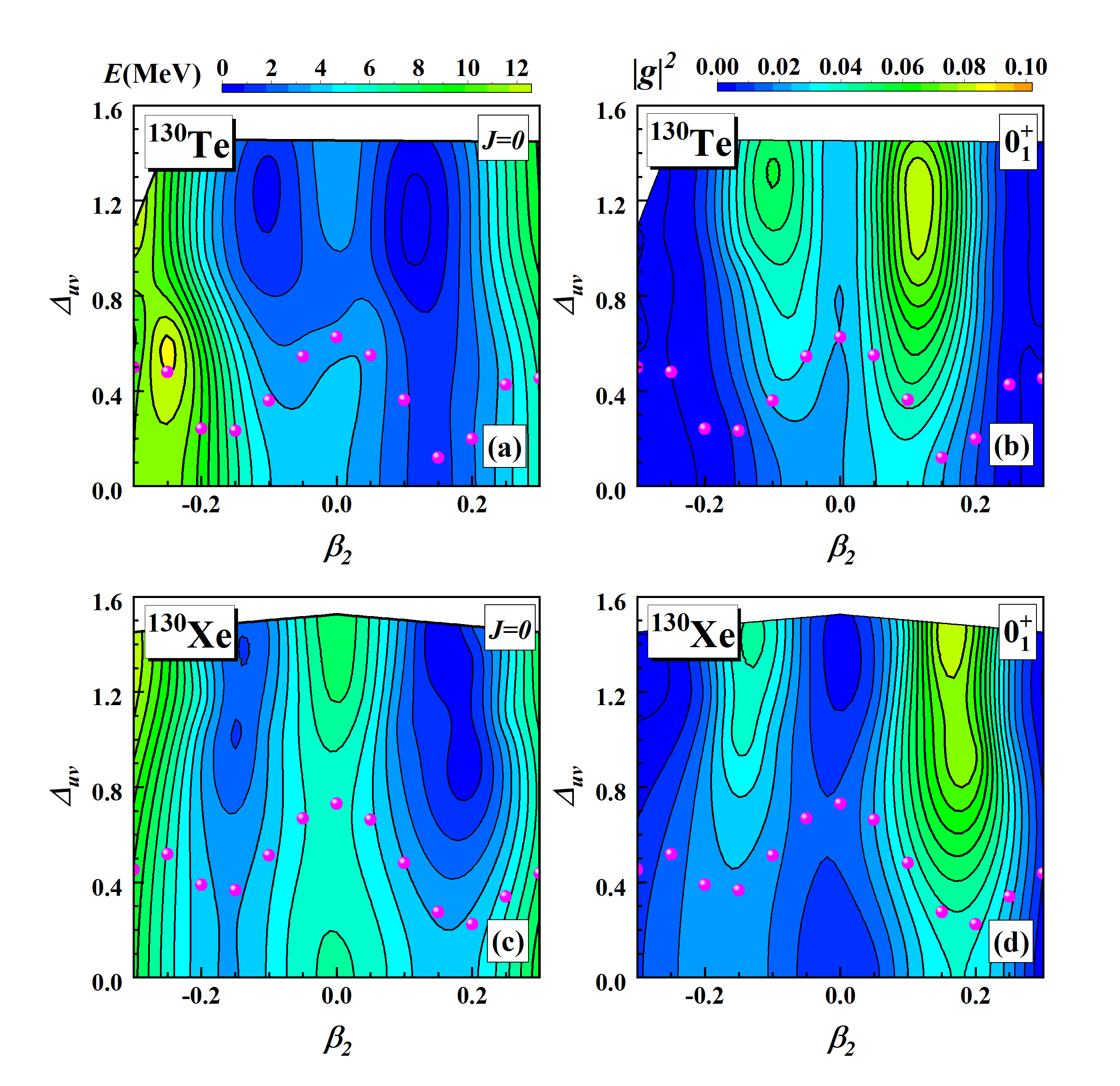} 
     \caption{The energies of states (normalized to the global energy minimum) for $^{130}$Te (a) and $^{130}$Xe  (c) with projection onto the particle numbers ($N, Z$) and angular momentum ($J=0$) as a function of $\beta_2$ and $\Delta_{u\varv}$. The distribution of the square of collective wave functions $|g^{0}_1|^2$ for the ground states of $^{130}$Te (b) and $^{130}$Xe (d) in the $\beta_2$-$\Delta_{u\varv}$ plane. The red dots are the same as those in Fig.~\ref{fig:MF_PES_TeXe}.}
     \label{fig:EJ0_wfs_TeXe}
 \end{figure}

Since the NMEs of $0\nu\beta\beta$ decay are sensitive to the isovector pairing strengths and also the pairing gaps of the mean-field configurations, it is natual to take pairing gap parameter $\Delta_{uv}$ as one of the generator coordinates in the GCM calculations. It provides a way to quantify the isovector pairing fluctuation effect on the  NMEs of $0\nu\beta\beta$ decay in the MR-CDFT calculation using the adjusted isovector pairing strengths. Fig.~\ref{fig:MF_PES_TeXe} shows the mean-field energy surfaces of \nuclide[130]{Te} and \nuclide[130]{Xe} in the $(\beta_2, \Delta_{u\varv})$ plane, where the average pairing gap $\Delta_{u\varv})$ has been defined in Eq.(\ref{eq:average_pairing_gap}). It is seen that the energy minimum of \nuclide[130]{Te} is located around the spherical shape, but very soft against the change of quadrupole deformation parameter $\beta_2$ and pairing gap parameter $\Delta_{u\varv}$.  In contrast, the global energy minimum for \nuclide[130]{Xe} locates around $\beta_2\simeq0.15$, which is also soft along the $\Delta_{u\varv}$ direction. It was pointed out in Ref.~\cite{Vaquero:2011PLB} that the softness of the energy surface occurs in the region with pure configurations, corresponding to the region with low level density. After the restoration of particle numbers and angular momentum with $J=0$, two energy minima competing in energy show up around $\beta_2=\pm0.10$ in  \nuclide[130]{Te}, as shown in Fig.~\ref{fig:EJ0_wfs_TeXe}. In particular, the two energy minima locate at the states with pairing gap parameter ($\Delta_{u\varv}=1.2$), larger than that ($\Delta_{u\varv}=0.8$) of the mean-field energy minimum, indicating that pairing correlation effect could be enhanced after considering pairing fluctuation. Indeed, the wave function $|g^{0}_1|^2$ of the ground state $(0^+_1)$, defined as
\begin{equation} 
g^{J}_{\sigma}(\mathbf{q})
=\sum_{\mathbf{q}'}  \left[{\cal N}^{J}_{00}(\mathbf{q},\mathbf{q}')\right]^{1/2}f^{J}_{\sigma}(\mathbf{q}'),
 \end{equation}  
 is concentrated around the two energy minima of projected states with large average pairing gaps. A similar phenomenon is also observed in \nuclide[130]{Xe}. Actually, it is understandable that the beyond mean-field effect arising from symmetry restoration generally deepens the symmetry-breaking states, generating a pronounced minimum as found in the near-spherical nuclei and triaxial $\gamma$-soft nuclei~\cite{Hayashi:1984PRL}.

\begin{figure}[]
	\centering{ \includegraphics[width=\columnwidth]{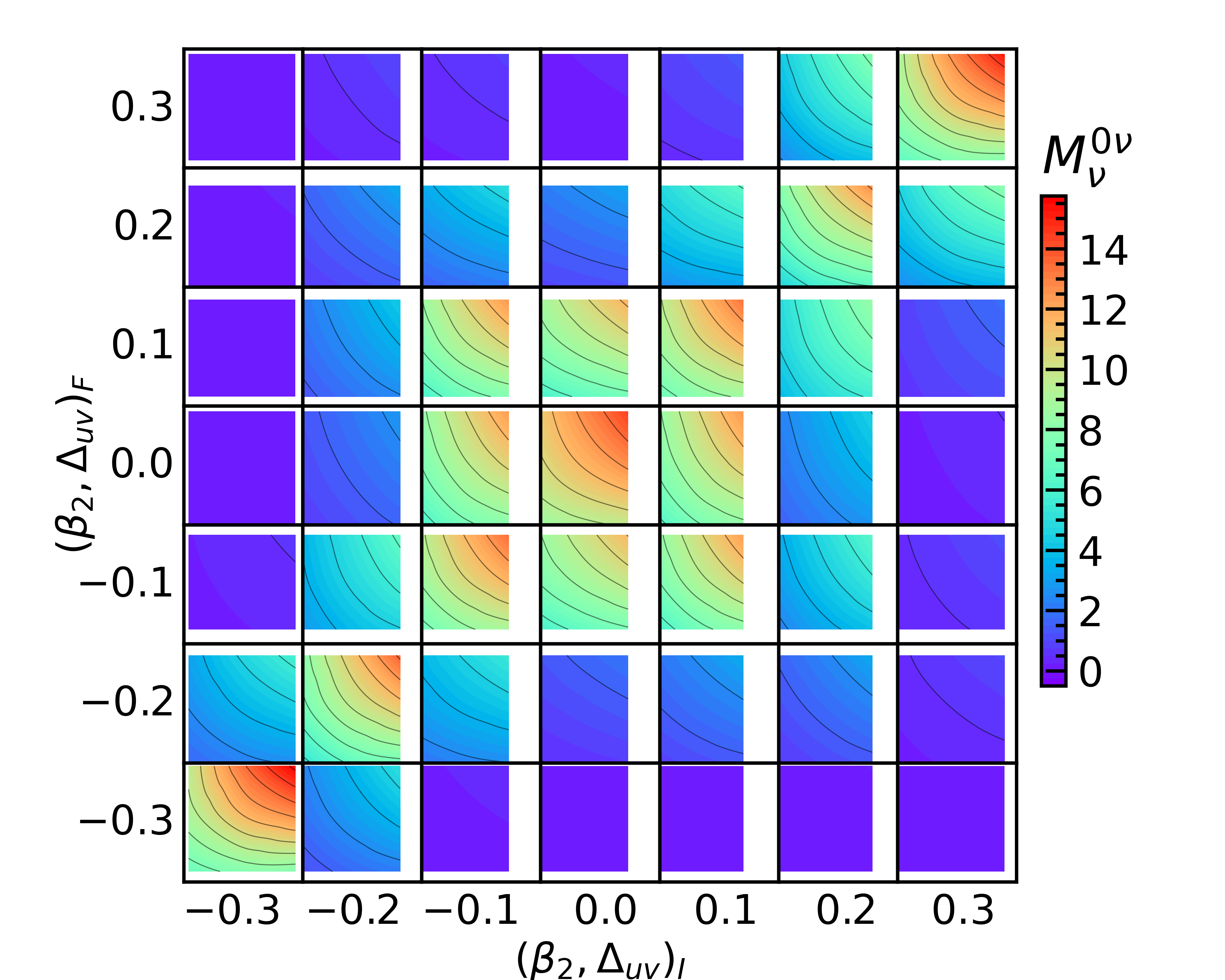}}  
	\caption{The normalized NME $\tilde M^{0\nu}_{\nu}(\mathbf{q}_F, \mathbf{q}_I)$ (\ref{eq:normalizedNME}) of $0\nu\beta\beta$ decay from the calculation with the exchange of light Majorana neutrinos as a function of the parameters $(\beta_2, \Delta_{u\varv})$ of $\nuclide[130]{Te}$ and $\nuclide[130]{Xe}$, where quadrupole deformation parameter $\beta_2$ in each subfigure is fixed to different value changing from $-0.30$ to $+0.30$, while the pairing gap $\Delta_{u\varv}$ varies from 1.0 MeV to 2.0 MeV.}
	\label{fig:NME_light_heavy_configuration_dependent}
\end{figure}

 Figure~\ref{fig:NME_light_heavy_configuration_dependent} displays the configuration-dependence of the normalized  NMEs $\tilde{M}^{0\nu}_\nu(\mathbf{q}_F, \mathbf{q}_I)$ defined in (\ref{eq:normalizedNME}) for \nuclide[130]{Te}.  It is clearly shown that the NME is large if the initial and final nuclei have the same collective coordinates, i.e.,  $\mathbf{q}_I=\mathbf{q}_F$. In particular, the NMEs of the subfigures corresponding to the configurations of near-spherical shapes for both nuclei are similar and generally larger than those of other subfigures.  Within each subfigure, the NME increases smoothly with the pairing gaps of initial and final nuclei. It is consistent with the conclusion in the previous studies~\cite{Rodriguez:2010,Vaquero:2013,Song:2014,Yoshinaga:2018} that the states of initial and final nuclei with a stronger pairing correlation produce a larger NME. The distributions of the NMEs in the case of exchanging light and heavy neutrinos are similar.

\begin{table*}[bt]
    \centering
    \renewcommand{\arraystretch}{1.0}
    \tabcolsep=5pt 
    \caption{Comparison of NMEs $M^{0\nu}_{\nu/N}$ of $0\nu\beta\beta$ decay for the five candidate nuclei from GCM calculations with the relativistic EDF PC-PK1 and non-relativistic EDF D1S, ab initio VS-IMSRG calculation, deformed QRPA calculations with the Brueckner G matrix and Skyrme EDF SkM*, interacting Boson model (IBM2) and interacting shell model (ISM) calculations. In the two GCM results,  upper and lower boundary values correspond to the results of calculations with and without isovector pairing fluctuation. In the VS-IMSRG results, upper and lower boundary values correspond to the results of calculations with and without the contact transition operator.  }
    \begin{tabular}{ccccccrcrr}
      \toprule
       & Isotopes &GCM(PC-PK1)($\chi$)& GCM(PC-PK1,$\beta_2$) &GCM(D1S)&VS-IMSRG &dQRPA(G)&dQRPA(SkM$^*$)&IBM2&ISM \\
      \multicolumn{2}{c}{} & This work &\cite{Song:2017}&\cite{Rodriguez:2010,Vaquero:2013}& \cite{Belley2021PRL,Belley:2023}&\cite{Fang:2018}&\cite{Mustonen:2013}&\cite{Barea:2009PRC,Barea:2015}&\cite{Menendez:2017fdf}\\
        \hline
       \multirow{5}{*}{$M^{0\nu}_{\nu}$} &\nuclide[76]{Ge}&[$2.37,6.34$]&$6.04$&[$4.60,5.55$]&$[2.14(9), -]$  &$3.12$&$5.09$&$5.46$&$3.07$\\
       &\nuclide[82]{Se}&[$2.46,5.68$]&$5.30$&[$4.22,4.67$]&$[1.24(5), -]$ & $2.86$&-&$4.41$&$2.90$\\
       &\nuclide[100]{Mo}&[$6.01,9.40$]&$6.48$&[$5.08,6.59$]&-&-&-& $3.73$&-\\
       &\nuclide[130]{Te}&[$1.89,6.02$]&$4.89$&[$5.13,6.40$]& $[-, 1.96(44)]$ & $2.90$&$1.37$&$4.06$&$2.96$\\
       &\nuclide[136]{Xe}&[$2.27,5.06$]&$4.24$&[$4.20,4.77$]& $[-, 1.49(41)]$ & $1.11$&$1.55$&-&$2.45$\\
        \hline
        \multirow{5}{*}{$M^{0\nu}_{N}$} &\nuclide[76]{Ge}&[$120.02,280.10$]&$209.1$&-&-& $187.3$&-&$104.0$&$188$\\
       &\nuclide[82]{Se}&[$134.58,273.54$]&$189.3$&-&-& $175.9$&-&$82.9$&$175$\\
       &\nuclide[100]{Mo}&[$289.17,444.59$]&$232.6$&-&-&-&-& $164.0$&-\\
       &\nuclide[130]{Te}&[$113.06,315.79$]&$193.8$&-&-& $191.4$&-&$91.8$&$210$\\
       &\nuclide[136]{Xe}&[$127.95,263.01$]&$166.3$&-&-& $66.9$&-&$72.6$&$167$\\
       \bottomrule
    \end{tabular}
    \label{tab:NME_comparison}
\end{table*}

\begin{figure}[bt]
     \centering \includegraphics[width=\columnwidth]{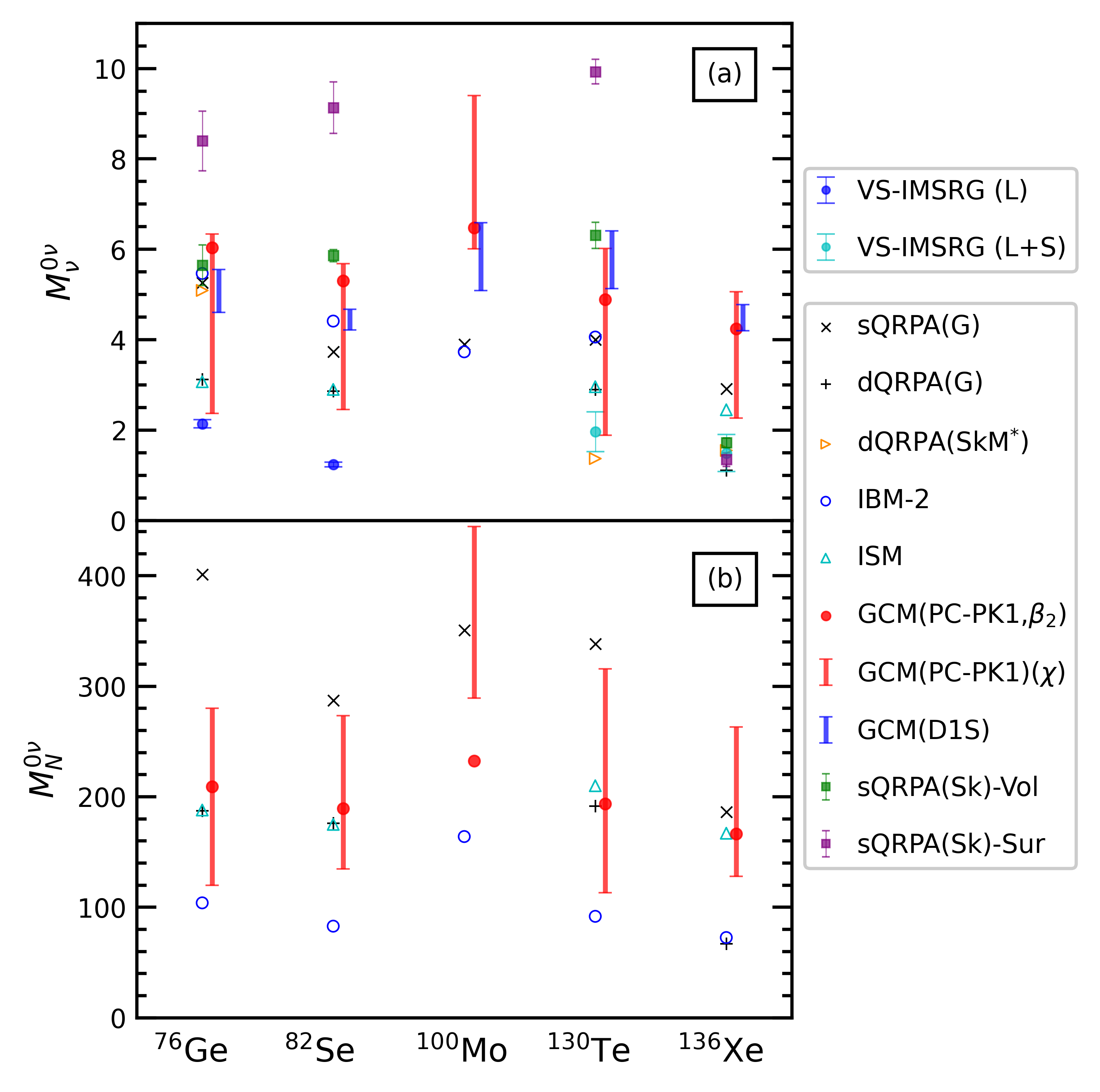} 
     \caption{The NMEs of $0\nu\beta\beta$ decay for the five candidate nuclei from the MR-CDFT calculation with pairing strengths adjusted to the excitation energies of $2^+_1$ states, labeled as GCM(PC-PK1)($\chi$). The results of calculations with and without the isovector pairing fluctuation are set as the upper and lower boundaries, respectively.  The NMEs from the GCM($\beta_2$) calculations using $\chi=1$~\cite{Song:2017} are indicated with red dots.  The NMEs are compared to those from other nuclear model calculations, including the GCM with the D1S force~\cite{Vaquero:2013}, interacting Boson model (IBM2)~\cite{Barea:2009PRC,Barea:2015}, ISM~\cite{Menendez:2017fdf}, the spherical~\cite{Hyvarinen:2015} and deformed QRPA~\cite{Fang:2018} based on the G matrix, the spherical~\cite{Lv:2023} and deformed~\cite{Mustonen:2013} QRPA based on Skryme EDFs, and the ab initio VS-IMSRG calculation with the long-range (L)~\cite{Belley2021PRL} and additionally  short-range (L+S)~\cite{Belley:2023} transition operators.}
     \label{fig:NME_comparison}  
\end{figure}

 The final NME $M^{0\nu}_{\alpha}$ is obtained from the configuration-dependent normalized NME $\tilde M^{0\nu}_{\alpha}(\mathbf{q}_F, \mathbf{q}_I)$  convoluted with the weight functions of the ground states for the initial and final nuclei from the solution of the HWG equation (\ref{eq:HWG}). After including the isovector pairing fluctuations, the total NMEs systematically increase, as shown in Tab.\ref{tab:NME_comparison}. Quantitatively, the NMEs of both $M^{0\nu}_\nu$ and $M^{0\nu}_N$ increase by a factor ranging from $56\%$ to $218\%$.  This enhancement effect from the isovector pairing fluctuation is consistent with the movements of the global energy minimum in the energy surface after symmetry restoration and the location of predominant configurations in the ground-state wave functions of both nuclei, as shown in Fig.~\ref{fig:EJ0_wfs_TeXe} for \nuclide[130]{Te}. Compared to the GCM($\beta_2$) calculation with only quadrupole shape fluctuation, the predominant configurations of the ground states for both nuclei in the GCM($\beta_2, \Delta_{uv}$) calculations are moved to regions with large average pairing gaps, resulting in a larger NME of $0\nu\beta\beta$ decay, see the upper boundary values of the GCM(PC-PK1)($\chi$) in Tab.\ref{tab:NME_comparison}. We note that the NMEs also increase with the isovector pairing fluctuation in the calculation based on the non-relativistic D1S force~\cite{Vaquero:2013}, showing  a similar enhancement pattern in the NMEs of the five candidates, even though the enhancement is much more moderate. The observed larger isovector pairing fluctuation effect in the GCM calculation based on the relativistic EDF  may be due to the lower level density around the Fermi energy resulting from a smaller effective nucleon mass in relativistic frameworks~\cite{Mughabghab:1998,Li:2018PPNP}. This leads to softer energy surfaces along the direction of $\Delta_{u\varv}$, as seen in Fig.~\ref{fig:MF_PES_TeXe} and Ref.~\cite{Vaquero:2011PLB}.

 The NMEs from different nuclear model calculations are compared in Fig.~\ref{fig:NME_comparison}. The detailed values, excluding those from the calculations without deformation effect, are presented in Tab.\ref{tab:NME_comparison}. The values of GCM calculations with and without the isovector pairing fluctuation are set as upper and lower boundaries, respectively.  It is seen from Fig.~\ref{fig:NME_comparison} that the NMEs from the MR-CDFT calculation with the mixing of only axially deformed shapes using the scaling factor of $\chi=1$ are generally located within the error bars. Previous Hamiltonian-based GCM studies have shown that including isoscalar pairing can reduce the NMEs~\cite{Hinohara:2014,Menendez:2016,Jiao:2017,Yao:2020PRL}. One may anticipate that the inclusion of isoscalar pairing in the EDF-based GCM calculations may also compensate partially the enhancement effect of isovector pairing fluctuation. Moreover, we have not considered the cranked states. On the other hand, however, the inclusion of cranked states in the GCM($\beta_2$)($\chi$)  calculation would lead to a somewhat larger scaling factor $\chi$ for the isovector pairing strengths, and thus a larger value of NME. As a result, the net effect of isoscalar pairing correlation and cranked states is expected to be moderate in the GCM($\beta_2, \Delta_{uv}$)($\chi$)  calculation. Of course, to draw a more solid conclusion on the effects of isoscalar pairing correlation and cranked states requires the extension of the MR-CDFT further, which is beyond the scope of this work. If we set the NMEs from the GCM($\beta_2$)($\chi$) and GCM($\beta_2, \Delta_{uv}$)($\chi$) calculations as the uncertainty of our calculations, the predicted NMEs are in line with most of other model calculations.  Extension of the present study with the effects of isoscalar pairing correlation~\cite{Hinohara:2014}, cranked states, and other possible higher-order deformed states~\cite{Yao:2015,Wang:2021} is expected to shrink this uncertainty.

\section{Summary}
\label{sec:summary}
We have explored the correlation relations among the NMEs of $0\nu\beta\beta$ decay, the excitation energies of $2^+_1$ and $4^+_1$ states, and the isovector pairing strengths within the multi-reference covariant density functional theory (MR-CDFT) with the mixing of axially-deformed shapes.  Based on the obtained correlation relations, we have adjusted the scaling factor of isovector pairing strengths for the five candidate nuclei to the excitation energies of $2^+_1$ states and computed the NMEs of $0\nu\beta\beta$ decay with and without the additional consideration of isovector pairing fluctuation. The results have shown that the description of the low-lying states has been improved with the adjusted pairing strengths in the MR-CDFT calculation. In the mean time, the predicted NMEs are reduced by about $12\%-62\%$. Furthermore, including the average isovector pairing gap as one additional generator coordinate in the MR-CDFT calculation provides a way to eliminate the dependence of the results on the choice of pairing gap parameter and it increases the NMEs by about $56\%-218\%$. The present study provides a promising starting point to determine the NMEs of $0\nu\beta\beta$ decay using the information of low-lying states within the MR-CDFT.

It is worth noting that the NMEs by the MR-CDFT with and without the  isovector pairing fluctuation effect can cause an uncertainty of a factor up to three, which is comparable to the observed discrepancy  among various nuclear models. It indicates that a precision determination of the NMEs of $0\nu\beta\beta$ decay with mean-field-based nuclear models demands a comprehensive consideration of pairing correlation  in atomic nuclei. The inclusion of additional isoscalar pairing correlation is not expected to change significantly the excitation energy of $2^+_1$ state, but is expected to reduce the impact of isovector pairing fluctuation and the NMEs of $0\nu\beta\beta$ decay. With the further extension of MR-CDFT by including additional isoscalar pairing correlation, cranked states and possible higher-order deformed states, and the pairing strengths constrained with the data of low-lying states, one can obtain NMEs with a greatly reduced uncertainty.

\section*{Acknowledgments}

 We thank J. Engel, D. L. Fang, H. Hergert,  C. F. Jiao,  G. Li, Y. F. Niu, and Y. K. Wang for fruitful discussions.  This work is partly supported by the National Natural Science Foundation of China (Grant Nos. 12141501, 11935003), Guangdong Basic and Applied Basic Research Foundation (2023A1515010936) and the Fundamental Research Funds for the Central Universities, Sun Yat-sen University.

 
%

\end{document}